\documentclass[10pt]{article}
\usepackage{amsfonts}
\usepackage{amssymb}
\usepackage{amsthm}
\usepackage{amsmath}
\usepackage{authblk} 
\usepackage{graphicx}
\usepackage{hyperref}
\usepackage{enumerate}
\usepackage{tikz}
\usetikzlibrary{decorations.pathmorphing, patterns,shapes}
\usepackage{color}
\usepackage[mathscr]{euscript}

\usepackage{soul,todonotes}
\usepackage{mathrsfs}
\usepackage{hyperref}
\usepackage{datetime}

\numberwithin{equation}{section}
\newcommand{\ee}{{\rm e}}
\newcommand{\x}{{\rm x}}

\newcommand{\dd}{{\rm d}}

\newcommand{\1}{1\!\!1}
\newcommand{\beq}{\begin{equation}}
\newcommand{\ene}{\end{equation}}

\theoremstyle{definition}

\newdateformat{daymonthyear}{\THEDAY \, \monthname[\THEMONTH] \THEYEAR}
\begin{document}

\title{Renormalisation in maximally symmetric spaces and semiclassical gravity in Anti-de Sitter spacetime}
\author{Benito A. Ju\'arez-Aubry$^{1}$\thanks{benito.juarezaubry@york.ac.uk} \, and Milton C. Mamani-Leqque$^{2,3}$\thanks{milton.mamani@unesp.br}}
\affil{$^1$ Department of Mathematics, University of York, Heslington, York YO10 5DD, United Kingdom }
\affil{$^2$ Universidad Nacional del Altiplano de Puno, Av. Floral 1153, 21001 Puno, Peru}
\affil{$^3$ Instituto de F\'isica Te\'orica, Universidade Estadual Paulista,
Rua Dr. Bento Teobaldo Ferraz, 271, 01140-070, S\~ao Paulo, S\~ao Paulo, Brazil}
\date{\daymonthyear\today}

\maketitle

\begin{abstract}
We obtain semiclassical gravity solutions in the Poincar\'e fundamental domain of $(3+1)$-dimensional Anti-de Sitter spacetime, PAdS$_4$, with a (massive or massless) Klein-Gordon field (with possibly non-trivial curvature coupling) with Dirichlet or Neumann boundary. Some results are explicitly and graphically presented for special values of the mass and curvature coupling (e.g. minimal or conformal coupling). In order to achieve this, we study in some generality how to perform the Hadamard renormalisation procedure for non-linear observables in maximally symmetric spacetimes in arbitrary dimensions, with emphasis on the stress-energy tensor. We show that, in this maximally symmetric setting, the Hadamard bi-distribution is invariant under the isometries of the spacetime, and can be seen as a `single-argument' distribution depending only on the geodesic distance, which significantly simplifies the Hadamard recursion relations and renormalisation computations.
\end{abstract}

\section{Introduction}
\label{sec:Intro}

Quantum field theory (QFT) in Anti-de Sitter spacetime (AdS) has gained substantial attention in the past years. This is undoubtedly in part because of its relevance in the AdS/CFT correspondence \cite{Maldacena:1997re} and holography, but also because AdS is interesting its own right. To mention two of its important features, first, AdS is a maximally symmetric spacetime, which allows one to put abstract techniques of QFT in curved spacetimes in a computationally accessible setting. Second, AdS is interesting because it is not a globally hyperbolic spacetime, but instead a good test-bed to see how techniques developed for QFT in globally hyperbolic spacetimes should be relaxed.

To mention some of the recent work, Dappiaggi and his collaborators have written a number of papers dealing with the construction of Klein-Gordon states with Robin boundary conditions in the Poincar\'e fundamental domain (PAdS) \cite{Dappiaggi:2016fwc} and in the universal cover (CAdS) \cite{Dappiaggi:2018xvw} of AdS. The case with dynamical Wentzell boundary conditions at the boundary of PAdS was studied in collaboration with one of us in \cite{Dappiaggi:2018pju, Dappiaggi:2022dwo}. Dynamical boundary conditions not only appear naturally in holographic renormalisation \cite{Skenderis:2002wp}, but are also central for the experimental verification of the dynamical Casimir effect \cite{Wilson:2011rsw}, as explained in \cite{Casimir1, Casimir2, Casimir3}.

Some rigorous results in the algebraic QFT framework appear in \cite{Dappiaggi:2017wvj}. Results on the propagation of singularities of Hadamard states that extend Radzikowski's microlocal spectrum condition in globally hyperbolic spacetimes \cite{Radzikowski} to asymptotically AdS spacetimes appear in \cite{Gannot:2018jkg, Dappiaggi:2020yxg}. Results from those papers allow for the construction of Hadamard states in asymptotically AdS spacetimes in \cite{Dappiaggi:2021wtr}. (See also the thesis of Marta \cite{Marta}.)

The behaviour of renormalised observables in AdS has been studied extensively in the literature by Winstanley and collaborators \cite{Kent:2014nya, Morley:2020ayr, Morley:2023exv, Thompson:2024vuj, Namasivayam:2022bky, Ambrus:2015mfa, Ambrus:2017cow}. The expectation value of the Klein-Gordon stress-energy tensor in CAdS$_{n}$ ($n = 2, \ldots 11$) is calculated in \cite{Kent:2014nya} with Neumann boundary conditions. The expectation value of the massless conformally coupled Klein-Gordon field squared with Robin boundary conditions at zero and finite temperature is studied in \cite{Morley:2020ayr}. Under the same conditions, \cite{Morley:2023exv} deals with the stress-energy tensor, and very recently \cite{Thompson:2024vuj} studies the back-reaction corrections to AdS spacetime.  The renormalised Klein-Gordon field squared with general mass and curvature coupling is studied in \cite{Namasivayam:2022bky}. Expectation values for the stress-energy tensor, current and field-squared for fermions are studied in \cite{Ambrus:2015mfa} and in \cite{Ambrus:2017cow} in the vacuum and in the case of finite temperature. 

The work of Pitelli and collaborators has emphasised the interplay between boundary conditions and spacetime symmetries. They have studied the field squared in \cite{Barroso:2019cwp} with hybrid Dirichlet-Robin boundary conditions. \cite{Pitelli:2019xeg} deals with the stress-energy tensor in PAdS$_2$ with Robin boundary conditions (see also \cite{Pitelli:2019pua, Pitelli:2021oil} for a study on the particle production in PAdS$_2$). The point is that states with generic Robin (or more general) boundary conditions are not invariant under the isometries of spacetime, unlike their Dirichlet or Neumann counterparts.

The construction of QFT in AdS is also important to understand QFT in relevant quotient spacetimes, such as the BTZ black hole in $n=3$ spacetime dimensions. Studies on semiclassical backreaction in BTZ appear in \cite{Casals:2016odj, Baake:2023gxx}.

In this paper, we are concerned with obtaining exact semiclassical gravity solutions in anti de-Sitter spacetime. More precisely, we shall work in the Poincar\'e fundamental domain of AdS in four spacetime dimensions, PAdS$_4$. Our interest stems from many angles. First of all, this is a natural extension to previous work by one of use finding semiclassical solutions in de Sitter spacetime \cite{PLB}, which are relevant to the cosmological constant problem, and to the related paper \cite{Gottschalk:2022bte}. Second, our paper serves as a proof of concept that semiclassical gravity can be defined in so-called globally hyperbolic spacetimes with timelike boundaries, in the sense of Ak\'e Hau, Flores and S\'anchez \cite{AkeFloresSanchez}. These are spacetimes where boundary conditions on the timelike boundary render wave equations well posed.

We are also motivated by two further questions, which we plan to address in the future. The first one has to do with the construction of semiclassical gravity solutions in asymptotically AdS spacetimes. A natural approach is to use Fefferman-Graham expansions, starting from semiclassical gravity in AdS at the lowest order, and then perturbatively construct the solutions near the timelike boundary. The second question has to do with whether semiclassical AdS has better or worse stability properties than its classical counterpart \cite{Bizon:2011gg, Bizon:2015pfa}.

As a way to achieve our goal of obtaining semiclassical gravity solutions in AdS, we study in some generality how to perform renormalisation, via Hadamard subtraction, in maximally symmetric spacetimes for the Klein-Gordon field with arbitrary non-negative mass and curvature coupling. We pay special attention to the stress-energy tensor, as this is the key observable appearing on the `right-hand side' of the semiclassical Einstein field equations. This is an addition to the literature in its own right, as it encompasses a number of situations of interest in a unified framework (see, for example, the literature cited above).

This paper is organised as follows. Section~\ref{sec:QFT} first provides a review of the Klein-Gordon theory in maximally symmetric spacetimes, and then gives a particularly useful representation for the Hadamard bi-distribution in this class of spacetimes, by showing that it is invariant under the spacetime isometries, and that it is a function of (half the square of) the geodesic distance only. Sec.~\ref{sec:TabMaxSym} is concerned with renormalisation in maximally symmetric spacetimes. Asymptotic expansions of the Hadamard coefficients are presented to sufficient accuracy to perform the renormalisation of the stress-energy tensor in $n=4$ spacetime dimensions. A simplified expression for the expectation value of the stress-energy tensor is given. The effects of changes in the renormalisation scale and the flat spacetime limit (as the radius of curvature tends to infinity) for the stress-energy tensor are studied. Sec.~\ref{sec:AdS} then applies the previous techniques to obtain, in closed form, the vacuum expectation value of the Klein-Gordon stress-energy tensor in PAdS$_4$ with Dirichlet and Neumann boundary conditions. Sec.~\ref{sec:Semiclass} then presents some semiclassical solutions in PAdS$_4$. Our final remarks appear in Sec.~\ref{sec:FR}.

\section{Quantum field theory in maximally symmetric spacetimes}
\label{sec:QFT}

This section provides an overview of maximally symmetric spacetimes with non-trivial curvature and of quantum states defined on these spacetimes. The main focus of the section is to show that the Hadamard condition for a free Klein-Gordon field adopts a particularly simple form, which makes renormalisation computationally economic. We achieve this by verifying that the Hadamard bi-distribution is invariant under the isometries of maximally symmetric spacetimes. It is clear that the simplifications that we find in this section for the Klein-Gordon field should apply to other theories defined by normally hyperbolic operators, but we do not discuss this any further and leave it as an interesting open question to address in the future.

\subsection{Maximally symmetric spacetimes}

Maximally symmetric spacetimes have positive, vanishing or negative constant curvature and, in $n$-dimensional spacetimes, $n(n+1)/2$ Killing vector fields generating isometries. The vanishing curvature case is Minkowski spacetime, with the Poincar\'e group as an isometry group. The positive curvature case is de Sitter spacetime, dS$_n$, with isometry group $O(1,n)$. The negative case is anti-de Sitter spacetime, AdS$_n$, with isometry group $O(2, n-1)$.

For maximally symmetric spacetimes, it is useful to express the curvature tensors in terms of their radius of curvature (see the discussion of Sec. \ref{subsec:PAdS} for details in the anti-de Sitter case). Setting $\rho^2 = \ell^2 >0$ for positive curvature and $\rho^2 = -\ell^2 <0$ for negative curvature, we have
\begin{align}
R_{abcd} &= \frac{1}{\rho^2}(g_{ac} g_{bd} - g_{ad} g{bc}), \\
R_{ab} & = \frac{n-1}{\rho^2} g_{ab}, \\
R & = \frac{n(n-1)}{\rho^2}.
\label{Curvature}
\end{align}

Maximally symmetric spacetimes solve the Einstein field equations, $R_{ab} - (1/2) R g_{ab} + \Lambda g_{ab} = 0$, with cosmological constant $\Lambda = (n-1)(n-2)/(2 \rho^2)$. More details about these spacetimes can be found in standard texts, such as \cite{HawkingEllis}. 

\subsection{Quantum fields and symmetric states}

We consider for concreteness a Klein-Gordon field, whose free algebra of observables in a globally hyperbolic spacetime (with or without timelike boundary), $\mathcal{M}=(M,g_{ab})$, is denoted by $\mathscr{A}(\mathcal{M})$, and well known to be the unital, $\star$-algebra generated by fields, $\Phi(f)$, smeared against test functions $f$ ($f \in C_0^\infty(M)$ if the boundary is empty), subject to relations
\begin{enumerate}
\item $\Phi(\alpha f + g) = \alpha \Phi(f) + \Phi(g)$ for $\alpha \in \mathbb{C}$ (linearity),
\item $\Phi(f)^\star = \Phi(\overline{f})$ (hermiticity),
\item $\Phi((\Box - m^2 - \xi R)f) = 0$ (Klein-Gordon equation) and
\item $[\Phi(f), \Phi(g)] = - i E(f,g) \1 $, where $E$ is the causal propagator of $\Box - m^2 - \xi R$ and $\1$ is the algebra unit (commutation relations).
\end{enumerate}

In the case of maximally symmetric spacetimes with non-trivial curvature, we focus on the case in which $\mathcal{M}$ is dS$_n$ or AdS$_n$. In the Anti-de Sitter case, one must impose boundary conditions on the spacetime boundary, and in the above construction different boundary conditions correspond to different test-function spaces and different causal propagators, $E$. See \cite{Dappiaggi:2017wvj} for some details in the Poincar\'e patch. While we will not dwell on details that are not central to this paper, in static spacetimes the definition of the observable algebra amounts to finding a suitable (non-unique) self-adjoint extension of a differential operator, which defines the spectral problem that is equivalent to solving the field equation, and considering test functions in an appropriate functional space.

Quantum (algebraic) states are linear maps $\omega: \mathscr{A}(\mathcal{M}) \to \mathbb{C}$ that are (i) normalised, i.e. $\omega (\1) = 1$, and (ii) positive, i.e. $\omega(A^\star A) \geq 0$ for any $A \in \mathscr{A}(\mathcal{M})$. The usual Hilbert space representation of the algebra can be obtained by means of the GNS construction, see e.g. \cite[Sec. 1.3]{KhavkineMoretti}.

Here, we consider locally Hadamard\footnote{Note that the local notion of the Hadamard condition is fully under control in globally hyperbolic domains of Anti-de Sitter spacetime, even if globally the wavefront set structure differs from the one characterised by Radzikowski, due to the presence of an asymptotic timelike boundary, see \cite{Gannot:2018jkg, Dappiaggi:2020yxg}.} quasi-free states of the Klein-Gordon field, which share the symmetries of the spacetime. It was noted in \cite{Allen, AllenJacobson} that the Wightman functions of symmetric states in a maximally symmetric spacetime are of the form
\begin{align}
G^+_\epsilon(\x, \x') = \mathcal{G}(\sigma_\epsilon(\x, \x')),
\label{Gsymm}
\end{align}
i.e., they can be seen as (single-argument) functions of Synge's world-function, $\sigma$, which measures half the squared geodesic distance between two points. More precisely, here $\sigma_\epsilon(\x, \x') = \sigma(\x, \x') + 2i[T(\x) - T(\x')]\epsilon + 2 \epsilon^2$ is a regularised version of $\sigma$, where $T$ is an arbitrary time function, which prescribes the distributional singular structure of $G_\epsilon$ (as $\epsilon \to 0^+$).

On the other hand, the Wightman function of a locally Hadamard state in a convex normal neighbourhood of spacetime takes the form
\begin{align}
G^+_\epsilon(\x, \x') = H_{\lambda \epsilon}(\x, \x') + W(\x, \x'),
\end{align}
where $H_{\lambda \epsilon}$ is a regularised version of the Hadamard bi-distribution, $H_\lambda$, 
\begin{align}
H_{\lambda}(\x, \x') & = \frac{\alpha_n}{2} \left( \Theta(n-5/2) \frac{u(\x, \x')}{\sigma^{n/2-1}(\x, \x')} + p(n) v(\x, \x') \log(\sigma(\x, \x')/\lambda^2) \right),
\label{eq:H}
\end{align}
with
\begin{align}
\alpha_n & = \begin{cases} 1/(2\pi), & \text{ if } n = 2; \\ \Gamma(n/2-1)/(2 \pi)^{n/2}, & \text{ if } n \geq 2;\end{cases} \\
p(n) & = \frac{1 + (-1)^{n+1}}{2}.
\end{align}

In Eq. \eqref{eq:H}, $u$ and $v$ are symmetric, smooth coefficients that can be (at least formally) defined through the Hadamard recursion relations, subject to appropriate boundary conditions on the diagonal, see e.g. \cite{Decanini-Folacci} for details. $\lambda$ is a fixed, arbitrary renormalisation scale. The factor of $5/2$ in the argument of the Heaviside $\Theta$ distribution in Eq. \eqref{eq:H} can be taken to be any real number in the open interval $(2,3)$.  

We shall now see that in maximally symmetric spacetimes, in analogy to Eq. \eqref{Gsymm}, we have
\begin{align}
H_\lambda(\x, \x') = \mathcal{H}_\lambda(\sigma(\x, \x')).
\label{Hsymm}
\end{align}

\subsection{The Hadamard condition in maximally symmetric spacetimes}
\label{subsec:HadExp}

In a convex normal neighbourhood of a globally hyperbolic spacetime region, the Hadamard bi-distribution, $H_\lambda$, admits the following expansion
\begin{align}
H_\lambda(\x, \x') & = \frac{\alpha_n}{2} \left[  \frac{\Theta(n-5/2) }{\sigma^{n/2-1}(\x, \x')} \sum_{m = 0}^{ \infty} u_m(\x, \x') \sigma^m(\x, \x') + p(n) \left( \sum_{m = 0}^N v_m(\x, \x')  \sigma^m(\x, \x') + O\left(\sigma^{N+1}\right) \right) \log\left(\frac{\sigma(\x, \x')}{\lambda^2}\right) \right],
\label{eq:Happendix}
\end{align}
where the order $N$ can be made arbitrarily large. The Hadamard recursion relations guarantee that, at any finite truncation of $H_\lambda$ of order $N$, say $H_\lambda^N$, the truncation will satisfy the Klein-Gordon equation modulo a $C^{N+1}$ term. Each coefficient in \eqref{eq:Happendix} can be written as \cite{Decanini-Folacci}
\begin{subequations}
\label{umvlExp}
\begin{align}
u_m(\x, \x') & = \sum_{j=0}^{J_m} \frac{(-1)^j}{j!}  u_{(mj) a_1 \ldots a_j}(\x) \sigma^{;a_1}(\x, \x') \ldots \sigma^{;a_j}(\x, \x') + O\left(\sigma^{(J_m+1)/2}(\x, \x')\right), \\
v_l(\x, \x') & = \sum_{k=0}^{K_l} \frac{(-1)^k}{k!}  v_{(lk) a_1 \ldots a_k}(\x) \sigma^{;a_1}(\x, \x') \ldots \sigma^{;a_k}(\x, \x') + O\left(\sigma^{(K_l+1)/2}(\x, \x')\right).
\end{align}
\end{subequations}

In practice, the order of expansions $I_m$ and $J_l$ in Eq. \eqref{umvlExp} can be chosen according to convenience in the renormalisation procedure. For example, for the renormalisation of the stress-energy tensor in $n =4$ dimensions, it suffices to set $J_0 = 4$, $K_0 = 2$ and $K_1 = 0$ to obtain an expansion of $H_\lambda$ with an error term of order $O(\sigma^{3/2})$ that does not contribute, in the limiting procedure of renormalisation, to the stress-energy tensor. (See Eq. \eqref{TabDef} below.)

In general, the coefficients $u_{(mj) a_1 \ldots a_j}$ and $v_{(lk)a_1 \ldots a_k}$ in Eq. \ref{umvlExp} are symmetric tensors, and depend only on the local geometry of spacetime and the field equation coefficients; they are functions of the metric tensor, the curvature and their derivatives,  and of $m$ and $\xi$. However, in maximally symmetric spacetimes, curvature is constant and does not depend on derivatives of the metric, cf. Eq. \eqref{Curvature}. Thus, the only available tensor indices for $u_{(mj) a_1 \ldots a_j}$ and $v_{(lk)a_1 \ldots a_k}$ are metric tensor indices. The general structure of the coefficients is hence that of symmetrised products of the metric tensor. It follows immediately that all coefficients with odd tensor rank vanish. For the coefficients with even tensor rank, we have the general structure
\begin{align}
u_{(mj) a_1 \ldots a_j} \sigma^{;a_1} \ldots \sigma^{;a_j} & = C_{mj} g_{(a_1 a_2} \ldots g_{a_{j-1} a_j)} \sigma^{;a_1} \ldots \sigma^{;a_j} = 2^{j/2} C_{mj} \sigma^{j/2},
\end{align}
where the $C_{mj}$ are constants, and similarly for the $v_{(lk)a_1 \ldots a_k}$ coefficients with constants $D_{lk}$. 

Hence, we can write Eq. \eqref{eq:Happendix} as
\begin{align}
H_\lambda(\x, \x') & = \frac{\alpha_n}{2} \left(  \frac{\Theta(n-5/2) }{\sigma^{n/2-1}(\x, \x')} \sum_{m = 0}^{\infty} U_m \sigma^m(\x, \x') + p(n) \left( \sum_{m = 0}^N V_m  \sigma^m(\x, \x') + O\left(\sigma^{N+1}\right) \right) \log(\sigma(\x, \x')/\lambda^2) \right),
\label{H:app2}
\end{align}
where the coefficients $U_m$ and $V_m$ are constants (i.e., spacetime independent), given by linear combinations of the $C_{rs}$ and of the $D_{rs}$ constants, respectively. Eq. \eqref{H:app2} shows that the Hadamard bi-distribution is invariant under the isometries of spacetime and of the form of Eq. \eqref{Hsymm}. The notation using capiatlised $U$ and $V$ emphasises that the $U_m$ and $V_m$ constants should not be confused with the diagonals, $u_m(\x, \x)$ and $v_m(\x, \x)$, of the coefficients appearing in Eq. \eqref{eq:Happendix}.

\subsection{The Hadamard recursion relations}
\label{subsec:HadamardRec}

The recursion relations for the coefficients $u_k$ and $v_k$ are remarkably simple. By Eq. \eqref{H:app2}, $u_k$ and $v_k$ can be seen as functions of the geodesic distance only. Furthermore, the van Vleck-Morette determinant in maximally symmetric spacetimes takes the form 
\begin{align}
\Delta = \left( \sqrt{-\frac{2 \sigma}{\rho^2}}  {\rm csch}  \left( \sqrt{-\frac{2 \sigma}{\rho^2}} \right) \right)^{n-1},
\label{DeltaClosed}
\end{align}
and defines also a (single-argument) function of $\sigma$.\footnote{See e.g. \cite{Kent:2014nya} for a derivation of Eq. \eqref{DeltaClosed} in the AdS case, for which a nice derivation of can be found in Kent's thesis \cite{KentThesis}.} In the remaining of this section, we shall sometimes use the abuse of notation $u_k(\sigma)$, $u_k(\sigma)$ and $\Delta(\sigma)$ and denote a derivative with respect to the argument by a prime.

Imposing, as usual, that
\begin{align}
(\Box_x -m^2 -\xi R(\x)) H_\lambda(\x, \x') + w(\x,\x') = 0,
\label{Hparametrix}
\end{align}
where $w$ is a smooth bi-function, yields the Hadamard recursion relations, which we now analyse.

\subsubsection{Spacetime dimension $n = 2$}

The recursion relations yield the tower of linear, non-homogeneous differetial equations
\begin{align}
4(k+1) \sigma v_{k+1}'(\sigma)  + 2(k+1)^2 v_{k+1}(\sigma)  - 2(k+1) \sigma \frac{ \Delta'(\sigma)}{\Delta(\sigma)}  v_{k+1}(\sigma) = - (\Box - m^2 - \xi R) v_k(\sigma),
\label{Dim2-v}
\end{align}
with the initial condition $v_0(\sigma) = - \Delta^{1/2}(\sigma)$. Since the cofficients of the differential equation \eqref{Dim2-v} are analytic, one can solve Eq. \eqref{Dim2-v} by a power series method writing
\begin{align}
v_k = \sum_{j = 0}^\infty v_{k j} \sigma^j,
\end{align}
with constant $v_{kj}$ coefficients, and using the series expressions
\begin{subequations}
\label{UsefulSeries}
\begin{align}
\Delta(\sigma) & = \sum_{j = 0}^\infty \frac{2^{j+1}(1-2^{2j-1}) B_{2j}}{(2 n)! (-\rho)^{2 j}} \sigma^j, \\
\frac{\Delta'(\sigma)}{\Delta(\sigma)} & = \frac{(n-1) }{2  \sigma } \left(1-\sqrt{-\frac{2 \sigma}{\rho^2}} \coth \left(\sqrt{-\frac{2 \sigma}{\rho^2}}\right)\right) = (1-n) \sum_{j = 1}^\infty \frac{2^{3j-1} B_{2j}}{(2j)! (-\rho)^{2j}} \sigma^{j-1} \\
\Box \sigma^k & = k(2k + n -2) \sigma^{k-1} -2 k \frac{ \Delta'}{\Delta} \sigma^k = k (2k+n-2) \sigma^{k-1} + k(n-1) \sum_{j = 1}^\infty \frac{2^{3j} B_{2 j}}{(2 j)! (-\rho)^{2 j}} \sigma^{k + j-1},
\end{align}
\end{subequations}
where $B_j$ is the $j$-th Bernoulli number. The recursion relations \eqref{Dim2-v} then yield algebraic solutions for the $v_{kj}$ constants recursively in $k$.

\subsubsection{Odd spacetime dimension $n \geq 3$}

In this case, the recursion relations yield the tower of differential equations
\begin{align}
2 (2k+4-n) \sigma u_{k+1}'(\sigma) + (k+1)(2k + 4-n) u_{k+1}(\sigma)  - (2k +4-n) \sigma  \frac{\Delta'(\sigma)}{\Delta(\sigma)}  u_{k+1}(\sigma)  = -(\Box - m^2 - \xi R) u_{k}(\sigma)
\end{align}
with the initial condition $u_0 = \Delta^{1/2}$.

Once again, the solutions can be obtained by the power series method, writing
\begin{align}
u_k(\sigma) = \sum_{j = 0}^\infty u_{k j} \sigma^j,
\end{align}
with constant $u_{kj}$ coefficients and using the series expressions \eqref{UsefulSeries}.

\subsubsection{Even dimension $n \geq 4$}

In this case, the recursion relations yield the tower of ordinary differential equations
\begin{align}
2 (2k+4-n) \sigma u_{k+1}'(\sigma) + (k+1)(2k + 4-n) u_{k+1}(\sigma)  - (2k +4-n)  \sigma \frac{\Delta'(\sigma)}{\Delta(\sigma)}  u_{k+1}(\sigma)  = -(\Box - m^2 - \xi R) u_{k}(\sigma)
\end{align}
for $k = 0, n/2-3$ (if $n > 4$) with the initial condition $u_0 = \Delta^{1/2}$ (if $n \geq 4$) and
\begin{align}
4(k+1) \sigma v_{k+1}'(\sigma)  + (k+1)(2k + n) v_{k+1}(\sigma)  - 2(k+1) \sigma \frac{ \Delta'(\sigma)}{\Delta(\sigma)}  v_{k+1}(\sigma) = - (\Box - m^2 - \xi R) v_k(\sigma),
\end{align}
with initial condition
\begin{align}
4 \sigma v_{0}'(\sigma) + (n-2) v_0(\sigma) - 2 \sigma  \frac{\Delta'(\sigma)}{\Delta(\sigma)} v_0(\sigma) = - (\Box - m^2 - \xi R) u_{n/2-2} (\sigma).
\end{align}

Once again, the solutions can be obtained by the power series method, writing
\begin{align}
u_k(\sigma) & = \sum_{j = 0}^\infty u_{k j} \sigma^j, \\
v_k(\sigma) & = \sum_{j = 0}^\infty v_{k j} \sigma^j,
\end{align}
with constant $u_{kj}$ and $v_{kj}$ coefficients, with the aid of the series expressions \eqref{UsefulSeries}.

\section{Renormalisation in maximally symmetric spacetimes}
\label{sec:TabMaxSym}

We set the spacetime dimension as $n = 4$, but it is clear how to extend the discussion to other dimensions. In this case, in a convex normal neighbourhood, the Hadamard bi-distribution \eqref{eq:H} takes the form
\begin{align}
H_\lambda(\x, \x') & = \frac{1}{2(2 \pi)^2} \left( \frac{\Delta^{1/2}(\x, \x')}{\sigma(\x, \x')} + v(\x, \x') \log(\sigma(\x, \x')/\lambda^2) \right),
\label{eq:H4}
\end{align}
where $\Delta$ is given by Eq. \eqref{DeltaClosed}. One can easily obtain from \eqref{DeltaClosed} the following expansion,
\begin{align}
\Delta^{1/2} % & = 1 + \frac{1}{12} R_{ab} \sigma^{;a} \sigma^{;b}-\frac{1}{24} R_{ab;c} \sigma^{;a}\sigma^{;b}\sigma^{;c} \nonumber \\
%& + \frac{1}{24} \left[\frac{3}{10} R_{(ab;cd)} + \frac{1}{12} R_{(ab}R_{cd)} + \frac{1}{15}R_{p(a\vert q \vert b} R^p{}_c{}^q{}_{d)}  \right] \sigma^{;a}\sigma^{;b}\sigma^{;c} \sigma^{;d} + O(\sigma^{5/2}).
& = 1 + \frac{\sigma}{2 \rho^2} + \frac{19 \sigma^2}{120 \rho^4} + O(\sigma^3).
\end{align}

Furthermore, if a state, $\omega$, is Hadamard and invariant under the isometries of the spacetime, its Wightman function takes the local form
\begin{align}
G^+_\epsilon(\x, \x') & = \mathcal{G}(\sigma_\epsilon(\x,\x')) = \frac{1}{2(2 \pi)^2} \lim_{\epsilon \to 0^+}  \left[  \frac{1}{\sigma_\epsilon(\x,\x')}\left(1 + \frac{\sigma(\x, \x')}{2 \rho^2} + \frac{19  \sigma^2(\x, \x')}{120 \rho^4} + O(\sigma^{3}(\x, \x'))\right) \right. \nonumber \\
& + \left( V_0 + V_1 \sigma(\x, \x') +O(\sigma^2) \right) \log(\sigma_\epsilon(\x, \x')/\lambda^2)  + w_0 + w_1 \sigma(\x,\x') + O(\sigma^{2}(\x,\x')) \Big],
\label{HadExpMaxSym}
\end{align}
where $V_0$ and $V_1$ are the state-independent constants, 
%\begin{align}
%V_0&= \frac{1}{2}m^2 + \left( \xi- \frac{1}{6} \right) \frac{1}{6 \rho^2}, \\
%V_1&= \frac{m^2}{8 \rho^2} + 3 \left(\xi - \frac{1}{6} \right) \frac{1}{\rho^4} + \frac{1}{60 \rho^4} + [v_1],
%\end{align}
%{\color{red}  CONTINUE HERE} 
see Sec. \ref{subsec:HadExp} above (and Eq. \eqref{GExpDiri} below for some explicit expressions in the negative curvature case), and $w_0$ and $w_1$ are free, state-dependent constants. 

Eq. \eqref{H:app2} justifies the the expansion of the singular structure of the two-point function in Eq. \eqref{HadExpMaxSym}. In order to justify the expansion for the smooth term in Eq. \eqref{HadExpMaxSym}, it suffices to remind oneself that the two-point function of symmetric states takes the form $G^+ = \mathcal{G}(\sigma)$. Thus, performing a power series in $\sigma$ and subtracting $H_\lambda$ (cf. Eq. \eqref{H:app2}) shows that the smooth part yields the Taylor series
\begin{align}
G^+(\x, \x')-H_\lambda(\x, \x') = \sum_{i = 0}^I w_i \sigma(\x, \x') + O\left(\sigma^{I+1}(\x, \x')\right),
\end{align}
with constant (i.e., spacetime independent) coefficients. 

We emphasise that, if the state is not symmetric, the coefficients $w_i$ will be in general replaced by smooth, symmetric bi-functions that are not invariant under the spacetime isometries. However, the structure of the covariant Taylor expansion of the Hadamard bi-distribution (and hence of the singular structure of Hadamard states) is still given by Eq. \eqref{H:app2}.

The expectation value of the renormalised stress-energy tensor is defined as usual by a point-splitting and Hadamard subtraction prescription,
\begin{subequations}
\label{TabDef}
\begin{align}
\omega (T_{ab}) &  := \lim_{\x' \to \x} \mathcal{T}_{ab} [G^+_0(\x,\x') - H_\lambda(\x, \x')] + \frac{1}{ (2 \pi)^2} g_{ab} [v_1] + \alpha_1 m^4 g_{ab} + \alpha_2 m^2 G_{ab} + \alpha_3 I_{ab} + \alpha_4 J_{ab} , \label{Stress-energyDef} \\
\mathcal{T}_{ab} &  := (1-2\xi ) g_{b}\,^{b'}\nabla_a \nabla_{b'} +\left(2\xi - \frac{1}{2}\right) g_{ab}g^{cd'} \nabla_c \nabla_{d'}  - \frac{1}{2} g_{ab} m^2 + 2\xi \Big[  - g_{a}\,^{a'} g_{b}\,^{b'} \nabla_{a'} \nabla_{b'} + g_{ab} g^{c d}\nabla_c \nabla_d + \frac{1}{2}G_{ab} \Big], \label{TabPointSplit2} \\
[v_1] & = \frac{1}{8} m^4 + \frac{1}{4} \left(\xi - \frac{1}{6}\right) m^2 R - \frac{1}{24}\left(\xi - \frac{1}{5} \right) \Box R   + \frac{1}{8} \left(\xi - \frac{1}{6} \right)^2 R^2 - \frac{1}{720} R_{ab} R^{ab} + \frac{1}{720} R_{abcd} R^{abcd}, \\
I_{ab} & := 2 R_{;ab} - 2 g_{ab} \Box R + \frac{1}{2}g_{ab} R^2 - 2 R R_{ab}, \label{Iab}, \\ 
J_{ab} & := R_{;ab} - \frac{1}{2} g_{ab} \Box R - \Box R_{ab} + \frac{1}{2} g_{ab} R^{cd} R_{cd} - 2 R^{cd} R_{cadb}.
 \label{Jab}
\end{align}
\end{subequations}

In maximally symmetric spacetimes with $n=4$, $I_{ab} = J_{ab} = 0$ \cite{PLB}, and one obtains the simple expression
\begin{align}
\omega(T_{ab}) & = \frac{1}{2(2\pi)^2} \left[ -w_1 + \frac{3 \xi}{\rho^2} w_0 - [v_1] \right]g_{ab} + \alpha_1 m^4 g_{ab} - \frac{3 \alpha_2 m^2}{\rho^2} g_{ab}, \label{TabFixedScale}
 \\
[v_1] & = \frac{m^4}{8} + \left(\xi - \frac{1}{6} \right) \frac{3 m^2}{\rho^2} + \left[18 \left(\xi - \frac{1}{6} \right)^2 - \frac{1}{60} \right] \frac{1}{\rho^4}, \label{v1}
\end{align}
(cf. Eq. (71) in \cite{Decanini-Folacci}). Note here that the coefficient $V_1$ in Eq. \eqref{HadExpMaxSym} is not the same as $[v_1] =  \lim_{\x' \to \x} v_1(\x, \x')$.

We note here that often a Wick-ordering notation, which emphasises the considered renormalisation procedure of the stress-energy tensor, is used in place of $\omega(T_{ab})$, whereby one would write explicitly  $\omega(:\! T_{ab} \!:_{H_\ell})$. This notation is especially accurate when the stress-energy tensor is considered as an element of the extended Wick algebra of observables of the field theory. We will write simply $\omega(T_{ab})$ throughout to ease the notation and because we think that no confusions can arise in the context of this paper.

\subsection{Changes of renormalisation scale}

Changes of the renormalisation scale introduce additional terms on the right-hand side of Eq. \eqref{TabFixedScale}. Set a new renormalisation scale as $\mu^{-1}$. Renormalising the stress-energy tensor with respect to this new scale adds to the right-hand side of Eq. \eqref{TabFixedScale} the constant term
\begin{align}
\lim_{\x' \to \x} \mathcal{T}_{ab}(H_\lambda(\x,\x') - H_{\mu^{-1}}(\x,\x')) & = -\frac{\log(\lambda^2 \mu^2)}{2(2\pi)^2}\lim_{\x' \to \x} \mathcal{T}_{ab}v(\x, \x') 
 = -\frac{\log(\lambda^2 \mu^2)}{2(2\pi)^2} \left[-V_1 + \frac{3 \xi}{\rho^2} V_0 \right]. % \nonumber \\
 %& = -\frac{\log(\lambda^2 \mu^2)}{2(2\pi)^2} \left[\frac{m^4}{8} + \frac{3}{2}\left( \xi - \frac{1}{6}\right) \frac{m^2}{\rho^2} \right] g_{ab}.
 \end{align}

\subsection{The flat spacetime limit}
\label{subsec:FlatLimit}

It was argued in \cite{PLB} that in the limit $\vert \rho^2 \vert \to \infty$ the expression for the stress-energy tensor ought to vanish, as required by Wald's stress-energy renormalisation axioms. Imposing this condition, we obtain
\begin{align}
\lim_{\vert \rho^2 \vert \to \infty} \omega(T_{ab}) = \frac{1}{2(2 \pi)^2} \left[ \lim_{\vert \rho^2 \vert \to \infty} \left(-w_1 + \frac{3 \xi }{\rho^2} w_0\right) -  \left( \log(\lambda^2 \mu^2) + 1\right) \frac{m^4}{8}\right] g_{ab} + \alpha_1 m^4 g_{ab} = 0.
\label{FlatSpaceLimit}
\end{align}

The limit of the state-dependent part in Eq. \eqref{FlatSpaceLimit} can be read off from the Minkowski two-point function, $G^+_{\rm M}$, for in the limit, the two-point function $G^+$ must satisfy the Klein-Gordon equation in flat spacetime. We have
\begin{align}
G^+_{\rm M} = \frac{m K_1\left(m \sqrt{2 \sigma_{\rm M} }\right)}{(2 \pi )^2 \sqrt{2 \sigma_{\rm M} }} = \frac{1}{2(2\pi)^2} \left[\frac{1}{\sigma_{\rm M}} + \left(\frac{m^2}{2} + \frac{m^4}{8} \sigma_{\rm M} + O(\sigma^2_{\rm M})\right)\log\left(\frac{m^2 \ee^{2 \gamma}\sigma}{2 \ee} \right) -\frac{3 m^4}{16}\sigma_{\rm M} \right] + O(\sigma_{\rm M}^2),
\end{align}
whereby we can identify $\lim_{\vert \rho^2 \vert \to \infty} w_0 = 0$ and $\lim_{\vert \rho^2 \vert \to \infty} w_1 = -3 m^4/16$.

In the massless case, the correct flat spacetime limit is satisfied. In the massive case, we obtain the relation
\begin{equation}
\alpha_1 = \frac{1}{16(2 \pi)^2}   \left( \log\left(\frac{\ee^{2 \gamma} \lambda^2 m^2}{2 \ee}\right) -2 \right).
\end{equation}

Finally, an expression for the expectation value of stress-energy tensor satisfying the correct flat spacetime limit is
\begin{align}
\omega(T_{ab}) & = \frac{1}{2(2\pi)^2} \left[ -w_1 + \frac{3 \xi}{\rho^2} w_0 \right]g_{ab}  - \frac{1}{2(2\pi)^2} \left[ 3 \left(\xi - \frac{1}{6} \right) \frac{m^2}{\rho^2} + \left(18 \left(\xi - \frac{1}{6} \right)^2 - \frac{1}{60} \right) \frac{1}{\rho^4} \right] g_{ab} %\nonumber \\ 
%& - \frac{3 \alpha_2 m^2}{\rho^2} g_{ab} -\frac{\log(\ell^2 \mu^2)}{2(2\pi)^2} \frac{3}{2}\left( \xi - \frac{1}{6}\right) \frac{m^2}{\rho^2} g_{ab}, 
+ \alpha_\mu(\xi) \frac{m^2}{\rho^2} g_{ab},
\label{TabWaldAxioms}
\end{align}
where we have defined $\alpha_\mu(\xi)$ as
\begin{align}
\alpha_\mu(\xi) = - 3 \alpha_2 -\frac{\log(\lambda^2 \mu^2)}{2(2\pi)^2} \frac{3}{2}\left( \xi - \frac{1}{6}\right).
\end{align}

\section{Quantum fields and semiclassical gravity in Anti-de Sitter spacetime}
\label{sec:AdS}

We are interested in finding solutions to the semiclassical gravity equations, cast in the form
\begin{align}
& G_{ab}(\x) + \Lambda g_{ab}(\x) = 8 \pi G_{\rm N} \omega(T_{ab}(\x)), \\
& (\Box - m^2 - \xi R(\x)) G^+(\x,\x') = (\Box' - m^2 - \xi R(\x') G^+ (\x,\x') = 0, \label{Semiclass2}
\end{align}
where $G^+$ is the Wightman function of the Klein-Gordon state $\omega$ in Anti-de Sitter spacetime. Here, the mass and curvature coupling of the field are allowed to take the values  $m^2 \geq 0$, $\xi \in \mathbb{R}$ and $\Lambda$ is a cosmological constant. We shall be focusing our attention on finding solutions in the Poincar\'e fundamental domain of Anti-de Sitter spacetime, PAdS$_4$, but some of the discussions in this section apply to PAdS$_n$ with arbitrary $n \geq 2$.

\subsection{The Poincar\'e fundamental domain of AdS spacetime}
\label{subsec:PAdS}

We begin by briefly introducing the Poincar\'e fundamental domain of Anti-de Sitter spacetime, PAdS$_n$. This spacetime can be seen as the `half' of AdS$_n$ covered by coordinates $t, x_i \in \mathbb{R}$ ($i = 1, \ldots, n-2$) and $z \in (0, \infty)$, whereby the spacetime line element takes a form conformal to the $n$-dimensional Minkowski half-space line element,
\begin{align}
\dd s^2 = %-\frac{(n-1)(n-2)}{2 \Lambda z^2} \left(-\dd t^2 + \dd z^2 + \sum_{i = 1}^{d-1} \dd x_i^2\right).
& = \frac{\ell^2}{z^2} \left(-\dd t^2 + \dd z^2 + \sum_{i = 1}^{n-2} \dd x_i^2\right).
\end{align}

Here $\ell$ is the radius of curvature of AdS$_n$ spacetime viewed as an embedded hyperboloid in an $(n+1)$-dimensional ambient, flat, pseudo-Riemannian manifold with metric signature $--+\ldots+$.\footnote{It is sometimes convenient to define the geodesic distance in PAdS$_n$, $s$, in terms of the chordal distance in the ambient manifold, $s_{\rm E}$, and the radius of curvature. We have the useful relation $\cosh\left(\frac{s}{\ell}\right) = 1+ \frac{s_{\rm E}^2}{\ell^2}$.} AdS$_n$ is a solution to the Einstein field equations with negative cosmological constant if the relation $\ell = \sqrt{-(n-1)(n-2)/(2 \Lambda)}$ holds between the radius of curvature, $\ell$, and the cosmological constant, $\Lambda$.

The asymptotic timelike boundary of PAdS$_{n}$ is approached as $z \to 0^+$, where boundary conditions must be prescribed for the matter fields defined in spacetime. We recall that the Riemann and Ricci tensors and the Ricci scalar in PAdS$_{d+1}$ take the simple form (cf. Eq. \eqref{Curvature})
\begin{subequations}
\begin{align}
R_{abcd} & = -\frac{1}{\ell^2}(g_{ac} g_{bd}- g_{ad}g_{bc}), \\ % =\frac{2 \Lambda}{(n-1)(n-2)} (g_{ac} g_{bd} - g_{ad}g_{bc}) = \frac{(n-1)(n-2)}{2 \Lambda z^4} (\eta^+_{ac} \eta^+_{bd} - \eta^+_{ad}\eta^+_{bc}), \\
R_{ab} & = -\frac{n-1}{\ell^2} g_{ab}, \\ % = \frac{2 \Lambda}{n-2} g_{ab} = - \frac{n-1}{z^2} \eta^+_{ab}, \\
R & = -\frac{n(n-1)}{\ell^2}. % = \frac{2 n \Lambda}{n-2}, 
\end{align}
\end{subequations}
%where $\eta^+_{ab}$ denotes the flat metric tensor in the Minkowski half-space.

\subsection{Quantum states in PAdS$_n$}

We consider a free Klein-Gordon field propagating in PAdS$_n$. As mentioned above, the details of the axiomatic quantisation of the theory appear in \cite{Dappiaggi:2017wvj}, and we shall not repeat them here. The crucial point is that PAdS$_n$ is a globally hyperbolic spacetime with timelike boundary in the sense of \cite{AkeFloresSanchez} and hence boundary conditions must be imposed together with the field equation.

For the purposes of this paper, we shall concentrate on Dirichlet and Neumann boundary conditions, as these preserve the isometries of the spacetime, see e.g. \cite{Pitelli:2019xeg}, and are therefore suitable for finding semiclassical Anti-de Sitter solutions.

The two-point functions of interest obey Eq. \eqref{Semiclass2} in a parameter space constrained by the Breitenlohner-Freedman bound, which we write in terms of a parameter, $\nu$,
\begin{align}
\nu := \frac{1}{2}\sqrt{1 + 4 \ell^2 \left[m^2 + \left(\xi - \frac{n-2}{4(n-1)} \right)R \right]} \geq 0.
\end{align}

The observation in \cite{Allen}, that for isometry-preserving states in maximally symmetric spacetimes the Wightman two-point function is a function of geodesic distance, allows one to obtain closed form expressions. To this end, it is convenient to define the function $u: \mathbb{R}_0^+ \to \mathbb{R}^+$ by
\begin{align}
u(\sigma) = \cosh^2\left( \sqrt{\sigma/(2 \ell^2)} \right),
\end{align}
where $\sigma = \frac{1}{2}s^2$ is Synge's worldfunction (half the squared geodesic distance). %Note that $u(\sigma) \geq 1$. 
We now quote results from \cite{Dappiaggi:2016fwc}.

\subsubsection{Dirichlet boundary conditions} Let $\nu \in (0, \infty)$. We call the Wightman two-point function with Dirichlet boundary conditions $G^+_{\rm (D)}$. It satisfies Eq. \eqref{Semiclass2} and $G^+_{\rm (D)}\vert_{z = 0} = G^+_{\rm (D)}\vert_{z' = 0} = 0$ and takes the form
\begin{align}
G^+_{\rm (D)}(x,x') = \lim_{\epsilon \to 0^+} \frac{\Gamma(\frac{n-2}{2}) \Gamma(\frac{1}{2}+\nu) \Gamma(\frac{3}{2}+\nu)}{2^{n} \pi^{\frac{n}{2}} \Gamma(1+2 \nu) \ell^{n-2}} u_\epsilon^{-\frac{n-1}{2}-\nu} {}_2F_1\left(\frac{n-1}{2}+\nu, \frac{1}{2} + \nu; 1 + 2 \nu; \frac{1}{u_\epsilon(x,x')} \right),
\end{align}
where $u_\epsilon$ is defined in terms of the regularised version of Synge's world function as $u_\epsilon = \cosh^2\left( \sqrt{\sigma_\epsilon/(2\ell^2)} \right)$.

\subsubsection{Neumann boundary conditions} Let $\nu \in (0,1)$. We call the Wightman two-point function with Neumann boundary conditions $G^+_{\rm (N)}$. It satisfies Eq. \eqref{Semiclass2} and $\partial_z G^+_{\rm (N)}\vert_{z = 0} = \partial_{z'}G^+_{\rm (N)}\vert_{z' = 0} = 0$ and takes the form
\begin{align}
G^+_{\rm (N)}(x,x') = \lim_{\epsilon \to 0^+} \frac{\Gamma(\frac{n-2}{2})\Gamma(\frac{1}{2}-\nu) \Gamma(\frac{3}{2}-\nu)}{2^{n} \pi^{\frac{n}{2}} \Gamma(1-2 \nu) \ell^{n-2} } u_\epsilon^{-\frac{n-1}{2}+\nu} {}_2F_1\left(\frac{n-1}{2}-\nu, \frac{1}{2} - \nu; 1 - 2 \nu; \frac{1}{u_\epsilon(x,x')} \right).
\end{align}

\subsection{The stress-energy tensor with Dirichlet and Neumann boundary conditions}

We henceforth focus on the case $n = 4$, with Breitenlohner-Freedman bound 
%$\nu = \frac{1}{2} \sqrt{1+4 m^2 \ell^2 - 48(\xi -1/6)} \geq 0$ and curvature radius $\ell = \sqrt{-3/\Lambda}$.
$\nu = \sqrt{9/4+\ell^2 m^2 -12 \xi} \geq 0$. 
 We give some details for the Dirichlet case. The Neumann case can be obtained by sending $\nu$ to its negative value. For the Dirichlet case, we have the covariant Taylor series
\begin{align}
& G^+_{\rm (D)} %&  = \frac{{\color{red} \Gamma(\frac{1}{2}+\nu) \Gamma(\frac{3}{2}+\nu)}}{16 \pi^{2} \Gamma(1+2 \nu) {\color{red}\ell^2} } u_\epsilon^{-\frac{3}{2}-\nu} {}_2F_1\left(\frac{3}{2}+\nu, \frac{1}{2} + \nu; 1 + 2 \nu; \frac{1}{u_\epsilon(x,x')} \right) 
%= \frac{1}{2(2 \pi)^2} \left\lbrace \frac{1}{\sigma}\left(1 - \frac{\sigma}{6 \ell^2} + \frac{\sigma^2}{10 \ell^4} + O(\sigma^3) \right) \right. \nonumber \\
%& + \left[ \frac{m^2}{2} + \frac{1}{\ell^2}\left(1 - 6 \xi \right) + \frac{1}{4 \ell^2 }\left(\ell^2 m^2-12 \xi +2\right) \left(\frac{m^2}{2}-\frac{6 \xi }{\ell^2}\right) \sigma + O(\sigma^2) \right] \left[ \log\left( \frac{\sigma}{2 \ell^2} \right) + \frac{\sigma }{6 \ell^2} \right] \nonumber \\
%& +  \left.  \frac{32 \left(4 \nu ^2-1\right) H_{\nu -\frac{1}{2}}-64 \nu ^2-48}{128 \ell^2}+\frac{ \left(4 \left(4 \nu ^2-9\right) \left(4 \nu ^2-1\right) H_{\nu -\frac{1}{2}}-80 \nu ^4+72 \nu ^2+51\right) \sigma}{256 \ell^4}    \right\rbrace \\
 = \frac{1}{2(2 \pi)^2} \left\lbrace \frac{1}{\sigma}\left(1 - \frac{\sigma}{2 \ell^2} + \frac{19\sigma^2}{120 \ell^4} + O(\sigma^3) \right) \right. \nonumber \\
& + \left[ \frac{m^2}{2} + \frac{1}{\ell^2}\left(1 - 6 \xi \right) + \frac{1}{4 \ell^2 }\left(\ell^2 m^2-12 \xi +2\right) \left(\frac{m^2}{2}-\frac{6 \xi }{\ell^2}\right) \sigma + O(\sigma^2) \right] \log\left( \frac{\sigma}{2 \ell^2} \right) \nonumber \\
& +  \left.  \frac{2 \left(4 \nu ^2-1\right) H_{\nu -\frac{1}{2}}-4 \nu ^2-\frac{1}{3}}{8 \ell^2}+\frac{ \left(4 \left(4 \nu ^2-9\right) \left(4 \nu ^2-1\right) H_{\nu -\frac{1}{2}}-80 \nu^4+ \frac{280}{3} \nu^2 + \frac{47}{5}\right) \sigma}{256 \ell^4}  +O(\sigma^2)  \right\rbrace,
\label{GExpDiri}
\end{align}
where $H_{\nu-1/2} = \psi(\nu +1/2)+\gamma$ is the analytic continuation of the harmonic number function, in terms of the digamma function, $\psi$. We can read off immediately
\begin{align}
w_0 & = \frac{2 \left(4 \nu ^2-1\right) H_{\nu -\frac{1}{2}}-4 \nu ^2-\frac{1}{3}}{8 \ell^2},  \\
w_1 & = \frac{ \left(4 \left(4 \nu ^2-9\right) \left(4 \nu ^2-1\right) H_{\nu -\frac{1}{2}}-80 \nu^4+ \frac{280}{3} \nu^2 + \frac{47}{5}\right)}{256 \ell^4}.
\end{align}

Using the results of Sec. \ref{sec:TabMaxSym}, we obtain, cf. Eq. \eqref{TabFixedScale}
\begin{align}
\omega_{\rm (D)}(T_{ab}) & = -\frac{1}{2(2\pi)^2} \left[ \frac{ \left(4 \left(4 \nu ^2-9\right) \left(4 \nu ^2-1\right) H_{\nu -\frac{1}{2}}-80 \nu^4 + \frac{280}{3} \nu^2 + \frac{47}{5}\right)}{256 \ell^4} \right. \nonumber \\ 
& \left. + \frac{3 \xi \left[2 \left(4 \nu ^2-1\right) H_{\nu -\frac{1}{2}}-4 \nu ^2-\frac{1}{3}\right]}{8 \ell^4} +  \frac{m^4}{8} - \left(\xi - \frac{1}{6} \right) \frac{3 m^2}{\ell^2} + \left(18 \left( \xi - \frac{1}{6} \right)^2 - \frac{1}{60} \right) \frac{1}{\ell^4} \right]g_{ab} \nonumber \\
& + \alpha_1 m^4 g_{ab} + \frac{3 \alpha_2 m^2}{\ell^2} g_{ab},
\label{TabDiri}
\end{align}
for the expectation value of the renormalised stress-energy tensor with Dirichlet boundary contions. A choice of the renormalisation constants with the correct flat-spacetime limit gives, cf. Eq. \eqref{TabWaldAxioms},
\begin{align}
\omega_{\rm (D)}(T_{ab}) & = -\frac{1}{2(2\pi)^2} \left[ \frac{ \left(4 \left(4 \nu ^2-9\right) \left(4 \nu ^2-1\right) H_{\nu -\frac{1}{2}}-80 \nu^4 + \frac{280}{3} \nu^2 + \frac{47}{5}\right)}{256 \ell^4} \right. \nonumber \\ 
& \left. +  \frac{3 \xi \left[2 \left(4 \nu ^2-1\right) H_{\nu -\frac{1}{2}}-4 \nu ^2-\frac{1}{3}\right]}{8 \ell^4} \right]g_{ab}  - \frac{1}{2(2\pi)^2} \left[ -3 \left(\xi - \frac{1}{6} \right) \frac{m^2}{\ell^2} + \left(18 \left(\xi - \frac{1}{6} \right)^2 - \frac{1}{60} \right) \frac{1}{\ell^4} \right] g_{ab} \nonumber \\
& - \alpha_\mu(\xi) \frac{m^2}{\ell^2} g_{ab}.
\end{align}

By a verbatim repeat, we obtain for Neumann boundary conditions
\begin{align}
\omega_{\rm (N)}(T_{ab})  & = -\frac{1}{2(2\pi)^2} \left[ \frac{ \left(4 \left(4 \nu ^2-9\right) \left(4 \nu ^2-1\right) H_{-\nu -\frac{1}{2}}-80 \nu^4 + \frac{280}{3} \nu^2 + \frac{47}{5}\right)}{256 \ell^4} \right. \nonumber \\ 
& \left. + \frac{3 \xi \left[2 \left(4 \nu ^2-1\right) H_{-\nu -\frac{1}{2}}-4 \nu ^2-\frac{1}{3}\right]}{8 \ell^4} +  \frac{m^4}{8} - \left(\xi - \frac{1}{6} \right) \frac{3 m^2}{\ell^2} + \left(18 \left( \xi - \frac{1}{6} \right)^2 - \frac{1}{60} \right) \frac{1}{\ell^4} \right]g_{ab} \nonumber \\
& + \alpha_1 m^4 g_{ab} + \frac{3 \alpha_2 m^2}{\ell^2} g_{ab}.
\label{TabNeu}
\end{align}

A choice of the renormalisation constants with the correct flat-spacetime limit gives, in the Neumann case,
\begin{align}
\omega_{\rm (N)}(T_{ab}) & = -\frac{1}{2(2\pi)^2} \left[ \frac{ \left(4 \left(4 \nu ^2-9\right) \left(4 \nu ^2-1\right) H_{-\nu -\frac{1}{2}}-80 \nu^4 + \frac{280}{3} \nu^2 + \frac{47}{5}\right)}{256 \ell^4} \right. \nonumber \\ 
& \left. +  \frac{3 \xi \left[2 \left(4 \nu ^2-1\right) H_{-\nu -\frac{1}{2}}-4 \nu ^2-\frac{1}{3}\right]}{8 \ell^4} \right]g_{ab}  - \frac{1}{2(2\pi)^2} \left[ -3 \left(\xi - \frac{1}{6} \right) \frac{m^2}{\ell^2} + \left(18 \left(\xi - \frac{1}{6} \right)^2 - \frac{1}{60} \right) \frac{1}{\ell^4} \right] g_{ab} \nonumber \\
& - \alpha_\mu(\xi) \frac{m^2}{\ell^2} g_{ab}.
\end{align}

We note that, because the stress-energy tensors \eqref{TabDiri} and \eqref{TabNeu} are proportional to the spacetime metric, they are completely determined by their trace, i.e., $\omega_{\rm (D/N)}(T_{ab}) = (1/4)\omega_{\rm (D/N)}(T_{c}{}^c) g_{ab}$ . In the conformally coupled case, $m^2 = 0$ and $\xi =1/6$, the trace contribution comes only from the trace anomaly and is state-independent. Thus, in this case, $\omega_{\rm (D)}(T_{ab})  = \omega_{\rm (N)}(T_{ab})$. As we shall see, the same equivalence between Dirichlet and Neumann holds for what we call here `effectively conformal' fields, which will be introduced below, for reasons that we will explain, although not owing to the trace anomaly in this case.

\section{Semiclassical gravity in PAdS$_4$}
\label{sec:Semiclass}

We present some solutions to the equations
\begin{align}
 \left( \frac{3}{\ell^2} + \Lambda\right) g_{ab}   = 8 \pi G_{\rm N} \omega_{\rm (D/N)} (T_{ab})
\label{sEFE}
\end{align}
in PAdS$_4$ with Dirichlet and Neumann boundary conditions. The solution space is characterised by the parameters of the theory, as the equations \eqref{sEFE} reduce to an algebraic relations due to the symmetries of the spacetime, as was observed in the de Sitter case \cite{PLB, Gottschalk:2022bte}.

If one does not demand that $\ell^2 = -3/\Lambda$ {\it a priori}, and keeps things as general as possible \cite{Gottschalk:2022bte}, one needs not impose the flat spacetime limit of Sec.~\ref{subsec:FlatLimit}. We therefore use Eq.~\eqref{TabDiri}/\eqref{TabNeu} for the right-hand side of Eq.~\eqref{sEFE}.

\subsection{Dirichlet boundary conditions}

The solutions to Eq. \eqref{sEFE} with Dirichlet boundary conditions are the solutions to the algebraic equation
\begin{align}
& \frac{1}{8 \pi G_{\rm N}}\left(\frac{3}{\ell^2} + \Lambda\right) - \alpha_1 m^4  - \frac{3 \alpha_2 m^2}{\ell^2}  \nonumber \\
& = -\frac{1}{2(2\pi)^2} \left[ \frac{ \left(4 \left(4 \nu ^2-9\right) \left(4 \nu ^2-1\right) H_{\nu -\frac{1}{2}}-80 \nu^4 + \frac{280}{3} \nu^2 + \frac{47}{5}\right)}{256 \ell^4} \right. \nonumber \\ 
& \left. + \frac{3 \xi \left[2 \left(4 \nu ^2-1\right) H_{\nu -\frac{1}{2}}-4 \nu ^2-\frac{1}{3}\right]}{8 \ell^4} + \frac{m^4}{8} - \left(\xi - \frac{1}{6} \right) \frac{3 m^2}{\ell^2} + \left[18 \left(\xi - \frac{1}{6} \right)^2 - \frac{1}{60} \right] \frac{1}{\ell^4} \right].
\label{SemiclassDirichlet}
\end{align}

Each solution is a point in the space of parameters $(\Lambda, \ell, m^2, \xi, \alpha_1, \alpha_2)$, with positive $\ell$, non-negative $m^2$,
\begin{align}
-\infty < \xi \leq \frac{4 \ell^2 m^2 + 9 }{48},
\label{xiInterval}
\end{align}
and real $\Lambda$, $\alpha_1$ and $\alpha_2$. 

Absorbing the parameters $\alpha_1$ and $\alpha_2$ into $\Lambda$ and the definition of $G_{\rm N}$, we can reduce the space of parameters to $(\Lambda, \ell, m^2, \xi)$. This means that the space of solutions is a subset of $\mathbb{R}^4$; for every admissible $m^2\geq 0$, $\xi$ obeying \eqref{xiInterval}, $\ell >0$, one can find a $\Lambda$ by imposing Eq. \eqref{SemiclassDirichlet}. 

In physical terms, Eq. \eqref{SemiclassDirichlet} constraints the space of parameters where semiclassical gravity solutions with strict Anti-de Sitter symmetry exist in the Poincar\'e fundamental domain. I.e., some AdS spacetimes admit semiclassical solutions with a scalar field with Dirichlet boundary conditions, while other do not, given a suitable fixing (or absorption) of renormalisation ambiguities.

Since there are {\it a priori} no physical reasons to constraint the parameters beyond the ranges that we have given, we proceed to study a few distinguished cases. Namely, the massless, and massive minimally coupled scenarios and the `effectively conformal' case, which includes the conformal coupling case. (The terminology for this last case is explained in Sec.~\ref{subsubsec:EffMf}.)

\subsubsection{The massless minimally coupled field}

\begin{figure}
\centering
\includegraphics[width=0.5\textwidth]{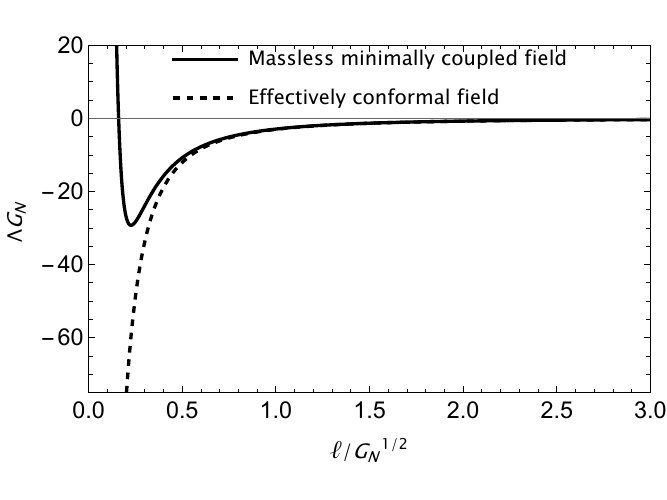}
\caption{The solid curve represents semiclassical PAdS$_4$ solutions with a massless and minimally coupled field with Dirichlet boundary conditions. The dashed curve represent semiclassical PAdS$_4$ solutions with a conformally coupled field with Dirichlet or Neumann boundary conditions.}
\label{Fig:R2}
\end{figure}

In this case, $m^2 = \xi = 0$ and $\nu = 3/2$. Each point along the solid curve of Fig. \ref{Fig:R2} represents a solution. The equations \eqref{SemiclassDirichlet} simplify to
\begin{align}
\Lambda = -\frac{3}{\ell^2}+\frac{29 G_{\rm N}}{120 \pi \ell^4}
\end{align}

It is easy to see that $\Lambda$ attains a minimum value when $\ell = \sqrt{29 G_{\rm N}/180}$, for which $\Lambda = -270 \pi/(29 G_{\rm N})$. As $\ell \to 0^+$, $\Lambda \to \infty$, and as $\ell \to \infty$, $\Lambda \to 0$, see Fig. \ref{Fig:R2}, 

\subsubsection{The massive minimally coupled field}

\begin{figure}
\centering
\includegraphics[width=0.5\textwidth]{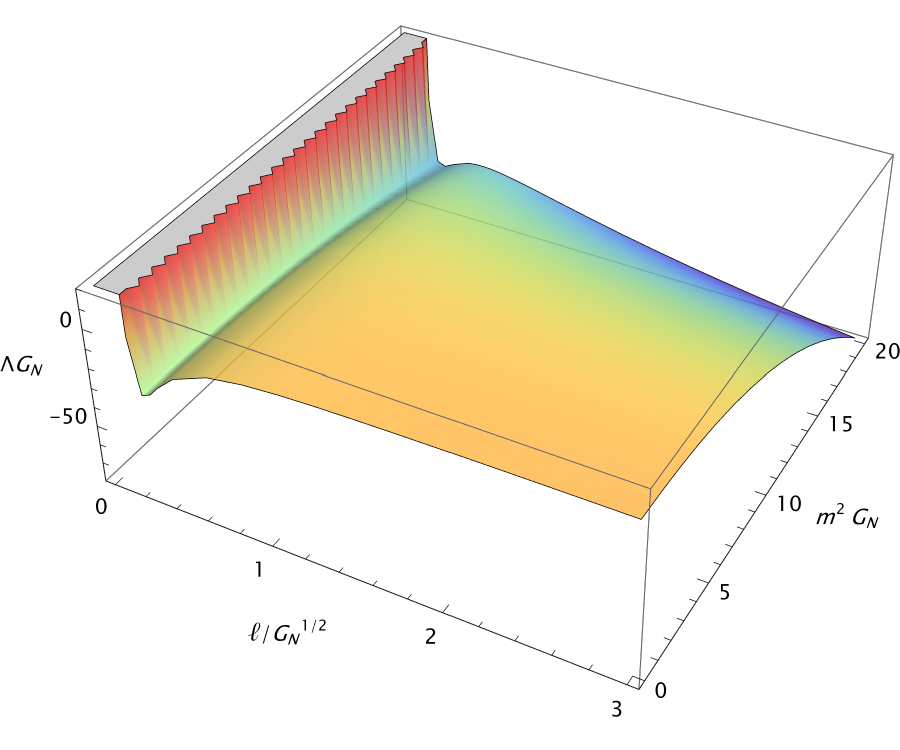}
\caption{Semiclassical PAdS$_4$ solutions with a massive and minimally coupled field with Dirichlet boundary conditions.}
\label{Fig:R3}
\end{figure}

In this case, $m^2 > 0$, $\xi = 0$ and $\nu = \sqrt{9/4 + \ell^2 m^2}$. The semiclassical equations do not have a simple closed form, but we sample numerically the space of solutions. Each point in the 3-dimensional plot of Fig. \ref{Fig:R3} represents a solution.

\subsubsection{Effectively conformal field}
\label{subsubsec:EffMf}

For the value $\nu = 1/2$, the Klein-Gordon equation in PAdS$_n$ takes the form of a massless Klein-Gordon equation in the Minkowski half-space {by a conformal transformation \cite{Dappiaggi:2016fwc}}. For $n=4$, this occurs when $\xi = (\ell^2 m^2 + 2)/12$. The semiclassical gravity equations simplify to
\begin{align}
\Lambda = -\frac{3}{l^2} -\frac{G_{\rm N} (1 - 5 \ell^2 m^2)}{120 \pi  l^4}
\end{align}

\begin{figure}
\centering
\includegraphics[width=0.5\textwidth]{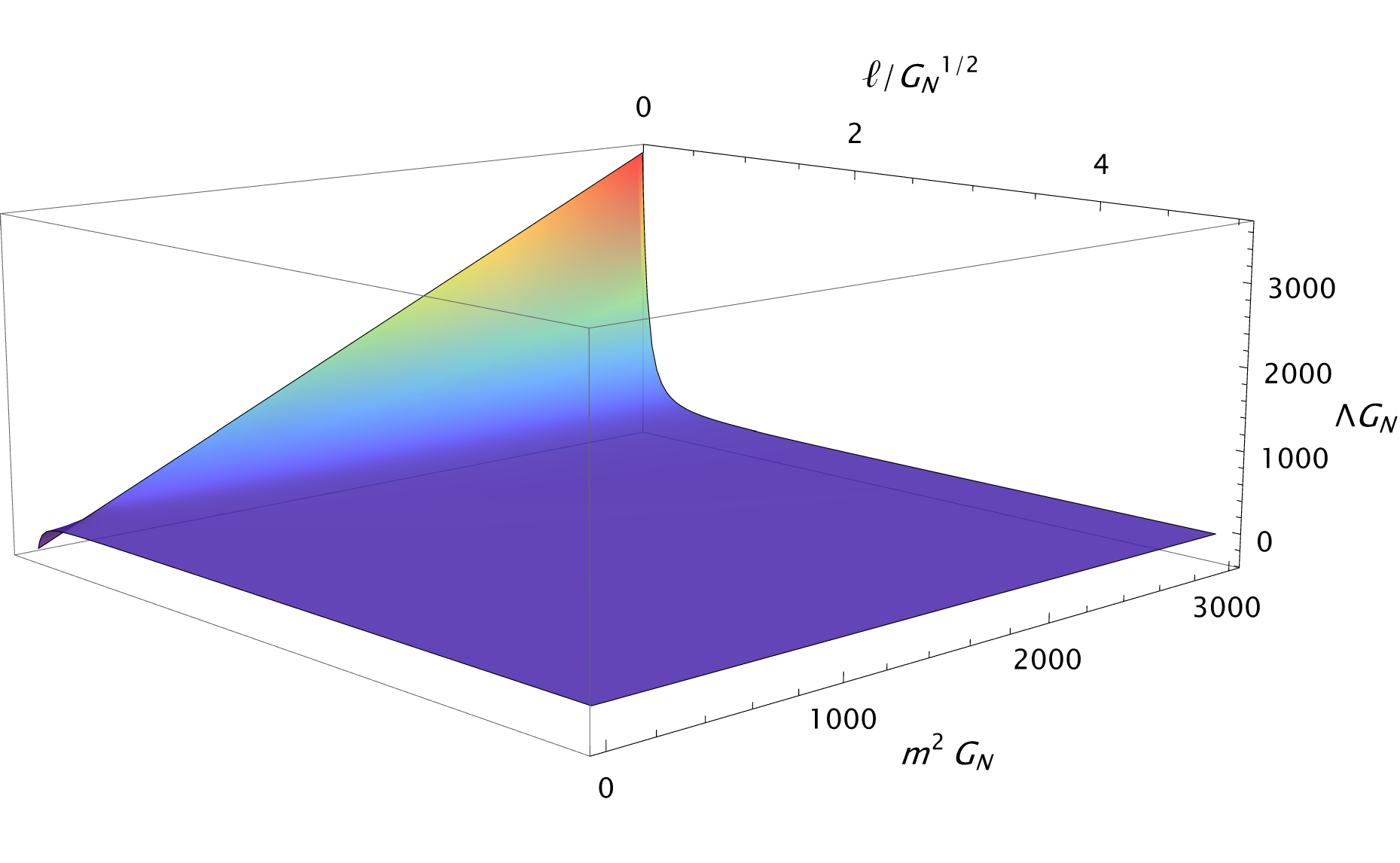}
\caption{Semiclassical PAdS$_4$ solutions with an effectively conformal field with Dirichlet or Neumann boundary conditions.}
\label{Fig:R4}
\end{figure}

It is straightforward to see that, for fixed $m^2$ as $\ell \to 0^+$, $\Lambda \to - \infty$, and as $\ell \to \infty$, $\Lambda \to 0$. At fixed $\ell$, the sign of $\Lambda$ depends on $m^2$. In particular, as $m \to \infty$, solutions exist even if $\Lambda >0$, with $\Lambda = 0$ whenever $m^2 = (5 G_{\rm N} \ell^2)^{-1}(3 \ell^2 + G_{\rm N})$. This should not alarm the reader, the Anti-de Sitter character of the solution is controlled by the radius of curvature of spacetime, $\rho^2 = -\ell^2$, which remains negative. The case $m^2 = 0$, $\xi = 1/6$ gives solutions for a conformal field with Dirichlet boundary conditions in PAdS$_4$. In this case, $\Lambda$, when viewed as a function of $\ell$, is a monotonically increasing, negative function. The solutions for the conformally coupled field are displayed along the dashed curve of Fig. \ref{Fig:R2}. The effectively conformal field solutions appear in Fig. \ref{Fig:R4}.

\subsection{Neumann boundary conditions}

The solutions to Eq. \eqref{sEFE} with Neumann boundary conditions are the solutions to the algebraic equation
\begin{align}
& \frac{1}{8 \pi G_{\rm N}}\left(\frac{3}{\ell^2} + \Lambda\right) - \alpha_1 m^4  - \frac{3 \alpha_2 m^2}{\ell^2} \nonumber \\
& = -\frac{1}{2(2\pi)^2} \left[ \frac{ \left(4 \left(4 \nu ^2-9\right) \left(4 \nu ^2-1\right) H_{-\nu -\frac{1}{2}}-80 \nu^4+72 \nu^2+ \frac{64}{3} m^2 \ell^2 -256 \xi + \frac{281}{3}\right)}{256 \ell^4} \right. \nonumber \\ 
& \left. + \frac{3 \xi \left[2 \left(4 \nu ^2-1\right) H_{-\nu -\frac{1}{2}}-4 \nu ^2-\frac{1}{3}\right]}{8 \ell^4} + \frac{m^4}{8} - \left(\xi - \frac{1}{6} \right) \frac{3 m^2}{\ell^2} + \left[18 \left(\xi - \frac{1}{6} \right)^2 - \frac{1}{60} \right] \frac{1}{\ell^4} \right],
\label{SemiclassNeumann}
\end{align}
which are points in the space of parameters $(\Lambda, \ell, m^2, \xi)$, with $\Lambda \in \mathbb{R}$ $\ell > 0$, $m^2 \geq 0$ and
\begin{align}
\frac{4 \ell^2 m^2 + 5 }{48}  < \xi < \frac{4 \ell^2 m^2 + 9 }{48}.
\label{xiNeu}
\end{align}

The physical interpretation of Eq. \eqref{SemiclassNeumann} is analogous to the Dirichlet case. It constraints the space of parameters where semiclassical gravity solutions with strict Anti-deSitter symmetry exist in the Poincar\'e fundamental domain, given a suitable fixing (or absorption) of renormalisation ambiguities.

Note that, in view of \eqref{xiNeu}, Neumann boundary conditions do not admit solutions with minimal coupling. In the case of the effectively conformal field, the Dirichlet and Neumann boundary conditions yield the same spacetime geometry, see Fig. \ref{Fig:R2}. This includes the case $m^2 = 0$, $\xi = 1/6$, which represents a conformal field with Neumann boundary conditions in PAdS$_4$.

In the case of the conformal coupling, the explanation for the equivalence between Dirichlet and Neumann boundary conditions relies in the fact that, as is well known, for conformal fields, the trace anomaly is state-independent, and depends only on the local geometry of spacetime. Taking the trace of Eq. \eqref{sEFE}, we see that the semiclassical equation is equivalent to
\begin{align}
\left( \frac{3}{\ell^2} + \Lambda\right)   = 2 \pi G_{\rm N} \omega_{\rm (D/N)} (T_{a}{}^a),
\end{align}
for which the right-hand side is independent of the Dirichlet or Neumann choice.

Furthermore, as we have mentioned above, in the effectively conformal case in PAdS$_4$ the Klein-Gordon equation is conformal to a massless field equation in Minkowski half-space. In this case, $m^2 + \xi R = R/6$ holds, even if $m^2 \neq 0$ and $\xi \neq 1/6$ . This means that the Klein-Gordon equation in PAdS$_4$  behaves as if the field were a conformally coupled field. The definition of the stress-energy tensor, however, is different to the conformally coupled case, because the parameters $m^2$ and $\xi$ take different values. In this case, the trace of the stress-energy tensor should not be seen as anomalous, since it will generically not vanish classically. Indeed, it is easy to verify that, for a classical configuration $\phi$, $T^a{}_a = (6\xi-1) \phi_{;a} \phi^{;a} + 6 \xi \phi \Box \phi -(\xi - 2 m^2) \phi^2$, which fails to vanish on-shell unless $m^2 = 0$ and $\xi = 1/6$ and will not vanish if merely $m^2 + \xi R = R/6$. However, it continues to hold that the Dirichlet and Neumann stress-energy tensors coincide in this case. This can be verified at the level of Eq. \eqref{TabDiri} and \eqref{TabNeu}.

\section{Final remarks}
\label{sec:FR}

This paper has dealt with the question of constructing semiclassical gravity exact solutions in Anti-de Sitter spacetime. We have focused our attention in particular to PAdS$_4$ with a Klein-Gordon field in the vacuum obeying Dirichlet or Neumann boundary conditions. Under a suitable re-definition of parameters, the solutions are characterised by a four-dimensional parameter space, including the mass term, $m^2$, curvature coupling, $\xi$, AdS radius, $\ell$, and the cosmological constant, $\Lambda$.

We present some solutions for the minimally coupled case and the `effectively conformal' case, which includes the conformal field. These solutions are points in the plots of Sec. \ref{sec:Semiclass}, which depict (2- and 3-dimensional) sections of the space of solutions (as subsets of the parameter space).

An interesting future direction is to extend this work to other field theories, such as the Dirac or Maxwell fields. In particular, it is interesting to see what are the appropriate boundary conditions that preserve the spacetime symmetries in these cases. It is also interesting to study the case of self-interacting field theories, extending the globally hyperbolic literature \cite{Hollands:2004yh, Costeri} to the case of spacetimes with timelike boundary, for which Anti-de Sitter spacetime is the simplest test-bed. For example, for a self-interacting scalar field, is it still the case that Dirichlet and Neumann boundary conditions respect the spacetime symmetries? We suspect that this will be the case, since, as we have seen, the Hadamard bi-distribution, which plays a crucial role in the renormalisation procedure, is invariant under the spacetime symmetries.

The situation is drastically different if one choses Robin boundary conditions. In the first place, the semiclassical Einstein equations do not take the simple form of Eq. \eqref{sEFE}. Instead, in this case, the lack of symmetry of the states breaks the maximal symmetry of spacetime and one should see corrections to the spacetime that leave the Anti-de Sitter class, for example, using semiclassical backreaction. As a result, there are generically no AdS semiclassical solutions for fields with Robin boundary conditions. Semiclassical corrections in this case give a deviation from the AdS class of spacetimes.

As we have seen, semiclassical gravity with what we have called here an effectively conformal field is insensitive to whether the boundary conditions are of Dirichlet or Neumann type, and one might wonder if this feature holds for the whole Robin class of boundary conditions.  The answer is in the negative. In fact, the two-point function with Robin boundary conditions for $\nu = 1/2$ is generically expressed as a mode sum which includes so-called bound state modes, which spoil the symmetry of the state \cite{Dappiaggi:2016fwc}. As a result, the expectation value of the stress-energy tensor is not proportional to the spacetime metric and cannot be fully characterised by the trace anomaly.

In this paper we have also given details on how to perform the renormalisation of non-linear observables in maximally symmetric spacetimes {\`a la} Hadamard in a computationally efficient way. The method relies on noting that the Hadamard bi-distribution shares the spacetime symmetries and takes a simplified form. This makes the computation of the Hadamard coefficients efficient and encompasses a number of situations that have been of interest in recent literature, as discussed in the Introduction. 

\section*{Acknowledgments}
BAJ-A is supported by EPSRC Open Fellowship EP/Y014510/1. At early stages, this work received support from CONAHCYT (formerly CONACYT), Mexico.

\vspace{5pt}

\noindent {\bf Rights Retention Statement:} For the purpose of open access, the corresponding author has applied a Creative Commons Attribution (CC BY) licence to any Author Accepted Manuscript version arising from this submission.

\end{document}